\documentclass[debug]{rmaa}

\usepackage{paralist}

\usepackage{psfrag,color}

\usepackage[latin1]{inputenc}

\usepackage{amsmath}

\title{An astrophysical perspective of life. The growth of complexity} 

\author{
  F. S\'anchez,\altaffilmark{1,2} 
  and E. Battaner,\altaffilmark{3}}

\altaffiltext{1}{Instituto de Astrof\'isica de Canarias, Spain.}

\altaffiltext{2}{Department of Astrophysics, ULL, Spain.}

\altaffiltext{3}{Instituto Carlos I de F\'isica Te\'orica y Computacional, Spain.}

\shortauthor{S\'anchez, \& Battaner}
\shorttitle{An astrophysical perspective of life}

\fulladdresses{
\item Francisco S\'anchez: Instituto de Astrof\'isica de Canarias, V\'ia L\'actea s/n, 38205 La Laguna, Tenerife, Spain \& Department of Astrophysics, University of La Laguna, Avda. Astrof\'isico Francisco S\'anchez s/n, 38206 Tenerife, Spain (fsm@iac.es).
\item Eduardo Battaner: Instituto Carlos I de F\'isica Te\'orica y Computacional, Universidad de Granada, Spain (battaner@ugr.es).}

\listofauthors{F. S\'anchez, \& E. Battaner}
\indexauthor{S\'anchez, F.}
\indexauthor{Battaner, E.}

\abstract{The existence of life is one of the most fundamental problems of astrophyisics. The intriguing existence of progressively complex and apparently improbable living beings should be a general tendency of life in the Universe. We are looking for general physical laws governing the growth of complexity in any astrophysical environment. We posit the existence of a vital scalar field. This scalar is sensitive to the gradient of the inverse of specific entropy, such that its distribution tends to very high values in the interior of living beings. Besides the classical mutations, vital field driven mutations only produce decrements of entropy. The field equations give rise to the existence of vital waves.  This theory is able to deal with both the origin of life and the evolution of life. We show that the growth of complexity is accelerated by the vital field.}

\resumen{La existencia de vida es uno de los problemas fundamentales de la astrofísica. La existencia intrigante de vivientes progresivamente complejos y aparentemente improbables debería ser una tendencia general de la vida en el Universo. Buscamos leyes físicas generales que gobiernen la emergencia de complejidad en cualquier sistema astrofísico. Proponemos la existencia de un campo vital escalar. Este escalar es sensible al gradiente del inverso de entropía específica, de forma que su distribución tiende a adquirir valores muy altos en el interior de los vivientes. Además de las mutaciones clásicas, hay otras mutaciones inducidas por el campo vital que sólo producen decrementos de entropía. Las ecuaciones del campo dan lugar a ondas vitales. Mostramos cómo el campo vital acelera el crecimiento de complejidad. Esta teoría puede aplicarse tanto el origen de la vida como a su evolución.}

\addkeyword{Hydrodynamics}
\addkeyword{Astrobiology}
\addkeyword{Cosmology}

\begin{document}
\maketitle

\section{Introduction}
   It is a fact that our Universe contains life, even if observations are still limited to only one planet. The origin and evolution of life in the Universe requires a general insight much beyond the precise knowledge that biology has reached in the particular case of living beings on Earth.
 
On the other hand, it is well known that, in the particular case of terrestrial life, a large impulse pushes living beings to perpetuate and benefit from any favourable circustance that arises and to progress. They tenaciously adapt themselves to the medium, evolving and even modifying this medium to improve their own evolution. Life efficiently uses any resource and any kind of energy independently of its origin. This well-known and fascinating behaviour makes it logical and necessary to  undertake the general problem of life in the evolution of the Universe as a fundamental astrophysical problem. 

In his celebrated book \emph{What is life?} \citet{schrodinger} wrote, ''From all that we have learnt about the structure of living matter, we must be prepared to find it working in a manner that cannot be reduced to the ordinary laws of physics''. This opinion of Schr\"odinger's inspired the present work, which aims to tackle essentially biological problems from a different viewpoint based on new physical laws. The complexity inherent in life and its evolution beckon us to try out new emergent models built from hitherto untried perspectives. 

We are not proposing an alternative to classical genetics but reinforce it with new physical laws. This theory is neither teleological nor is based on anthropic principle.

Many works have dealt with the physical interpretation of life, entropy, complexity and evolution \citep[some representative classic examples could be: ][]{prigogine81,prigogine84,maturana,mcshea,kauffman,margulis,schneider,krakauer,grassberger,day,gould,davies,adamsky,zeravcic} but their basic principles and scopes are  very different to the one presented here. There are many genetic and astrobiological topics dealing with the evolution of life in the Universe, but here we restrict ourselves to the entropic problem. This paper aims to provide a fresh insight from a purely physics-based point of view. 

We propose general physical laws for any kind of life in the Universe, terrestrial life just being a particular case. We need, however, to make a small number of general assumptions that are evident in terrestrial genetics; namely, that the existence of a molecular code similar or dissimilar to our DNA, that this code must be highly stable to subsist against replication and thermal noise, and that this stability is not completely perfect, so that it can be on rare occasions violated, thereby permitting mutations. The mechanism of natural selection should also be at work.

We introduce a vital field characterized by vital density and velocity. The vital field permeates the Universe and flows in space and time. We first define what this vital field \emph{is} and later consider what this vital field \emph{does}. 

The properties of the vital field are defined by two proposed equations bearing formal similarity with the basic equations of the Eulerian Fluid Mechanics. From these equations, we deduce the existence of vital waves. We then propose that vital density has at least two effects on the evolution of species. First, it induces another type of mutation characterized by decrements of specific entropy (called VID mutations, Vital Intensity Driven mutations). Second, vital waves enable a mechanism that speeds up the dissemination of mutations. 

The novel hypothesis proposed here may render some evolution processes more effective. With both VID mutations and standard mutations, natural selection continues to be the mechanism that decides the progress of evolution. We could even suggest that the new hypothesis provides support to the traditional theory of evolution and can therefore be considered as complementary to it. We are not at all proposing an alternative to the standard theory of evolution. We do not exclude standard genetical concepts; instead we simply add new ones, albeit based on new physics.
   
 \section{Mathematical derivations}
 
 \subsection{Vital field equations}
 
 The physical magnitude characterizing the complexity referred to here is specific entropy (entropy per unit mass, or entropy per baryon), $s$. For living beings we also use the thermodynamic concept of entropy, $S$, or rather its statistical mechanical formulation according to the interpretation of Boltzmann. We shall also speak of the entropy of DNA, or any other equivalent code, on the basis of the relation between entropy and information by taking into account the information contained microscopically in the genome. This entropy is based on the thermodynamic weight (number of microstates of a macrostate). Since the two should be intimately interrelated, there is no risk of misinterpretation. 

The scalar physical magnitude that defines the vital field is vital density, $\rho(x,y,z,t)$, defined for all space and time. The characteristics of the vital field are determined by two relations: the equation of vital flux and the continuity equation. 

We define the vital flux by the product $\rho \vec{v}$, where $\vec{v}$ is the characteristic fluid velocity of the spatial variations of vital density (not the velocity of propagation of the vitons, as we shall see later).

We write the equation of vital flux, i.e. using a hydrodynamic approach, as follows:
\begin{equation}
    {{\partial} \over {\partial t}} \left(\rho \vec{v} \right) + k_1 \nabla \rho - k_2 \nabla {1 \over s} = 0
    \label{eq1}
\end{equation}
where $k_1$ and $k_2$ are constants.
   
In order to interpret this equation, let us first suppose the simple case of a stationary state:
 \begin{equation}
   \nabla \left(k_1 \rho - {k_2 \over s} \right) = 0
   \label{eq2}
 \end{equation}

The simplest way to integrate this equation is by ignoring the sign of the gradient. We then obtain an equation that we can work with:
 \begin{equation}
   \rho \propto {1 \over s}
   \label{eq3}
 \end{equation}

In other words, only for stationary states, those points corresponding to the interior of a living being with very low entropy will have high vital density; outside, the density will be very low. The density will not be homogeneous inside because the constituent organs will not possess the same complexity, i.e. the same entropy.  

In the general case around and within a living being, without considering steady state conditions, Eq.~\ref{eq1} tells us that the vital flux varies, owing to two gradients of opposing tendencies. The gradient of the inverse of the specific entropy points from the exterior to the interior and will tend to generate a vital flux towards the interior of the living being. The gradient of density also points inwards, but because of its negative sign, generates a vital flux outwards. 

If the density distribution tends to have very high values inside, such values cannot be attained instantaneously. If, for example, the living being is displaced, the density distribution 'follows', although with a slight delay that is inappreciable if the velocity of the living being is very much less than the speed of light. For that to occur it is necessary that the constants $k_1$ and $k_2$ be very high (close to $c^2$). In the case of the displacement of a living being, both gradients act jointly to make the distribution of vital density and the inverse of the specific entropy coincide. This implies that, in practice, the density distribution seems to be in step with, or frozen into, the body of the living being.

Vital density does not represent life complexity, complexity being represented by low specific entropy. But low entropy attracts the density as shown by Eq.~\ref{eq1} and this effect takes place over very short characteristic times. Steady state conditions can be quickly reached and then the distribution of both the density and the inverse of specific entropy may coincide. 

The causes of variability in vital flux correspond to rapid processes with very short characteristic times, in contradistinction to long-duration processes (i.e. those that operate on evolutionary time scales), which we shall discuss further on. 

We call the second of the vital field equations the equation of continuity. In fluid mechanics the equation of continuity establishes the conservation of mass; here, in a similar way, the integrated density, is conserved throughout space from -$\infty$ to +$\infty$. We therefore have:
 \begin{equation}
   \iiint_{-\infty}^{+\infty} \rho d\tau = constant
   \label{eq4}
 \end{equation}
where d$\tau$ is the element of volume. The triple integral in this formula means that the total integrated density of the vital field in the whole Universe remains constant in time.

The equation of continuity tells us that if, in an element of volume, the divergence of the flux is positive (i.e. if more flux enters than leaves the volume element), the density will diminish:
\begin{equation}
  {{\partial \rho} \over {\partial t}} = - \nabla \cdot \left(\rho \vec{v} \right) 
  \label{eq5}
\end{equation}

This equation establishes the conservation of interated density. Interpretation of the equation of continuity of density brings us to the physics that we wish to highlight. During its lifetime a living being is a stable system. The only great changes occur at the foetal stage, infancy, and death. Throughout the rest of its life the organism remains extraordinarily constant. If its alimentation is of a chemical nature, the process of ingestation introduces entropy. As the system is in equilibrium, this increase of entropy must be eliminated through the expulsion of mass in the form of faeces and urine, in the case of terrestrial warm-blooded animals the process of perspiration results in heat exchange with the environment through the latent heat of vaporization, and so on.  A living organism may be thought of as a stable heat engine (or a cyclic engine with a period of the order of one day). A stable living organism is a great ejector of entropy into its immediate surroundings. The ejection of entropy to the surroundings should even be inherent to the very concept and definition of life. 

For any given instant of time in the Universe the density distribution associated with a living emitity will be very high (even though non-uniformly distributed in the interior of the organized living entity) and very low in its immediate surroundings, asymptotically approaching the value associated with lifelessness towards infinity. The mean value would correspond to the lifeless value; the high value of density in the interior of the organism would compensate for the minimum value in its immediate surroundings through the ejection of entropy. 

\begin{figure}
   \centering
   \includegraphics[width=\hsize]{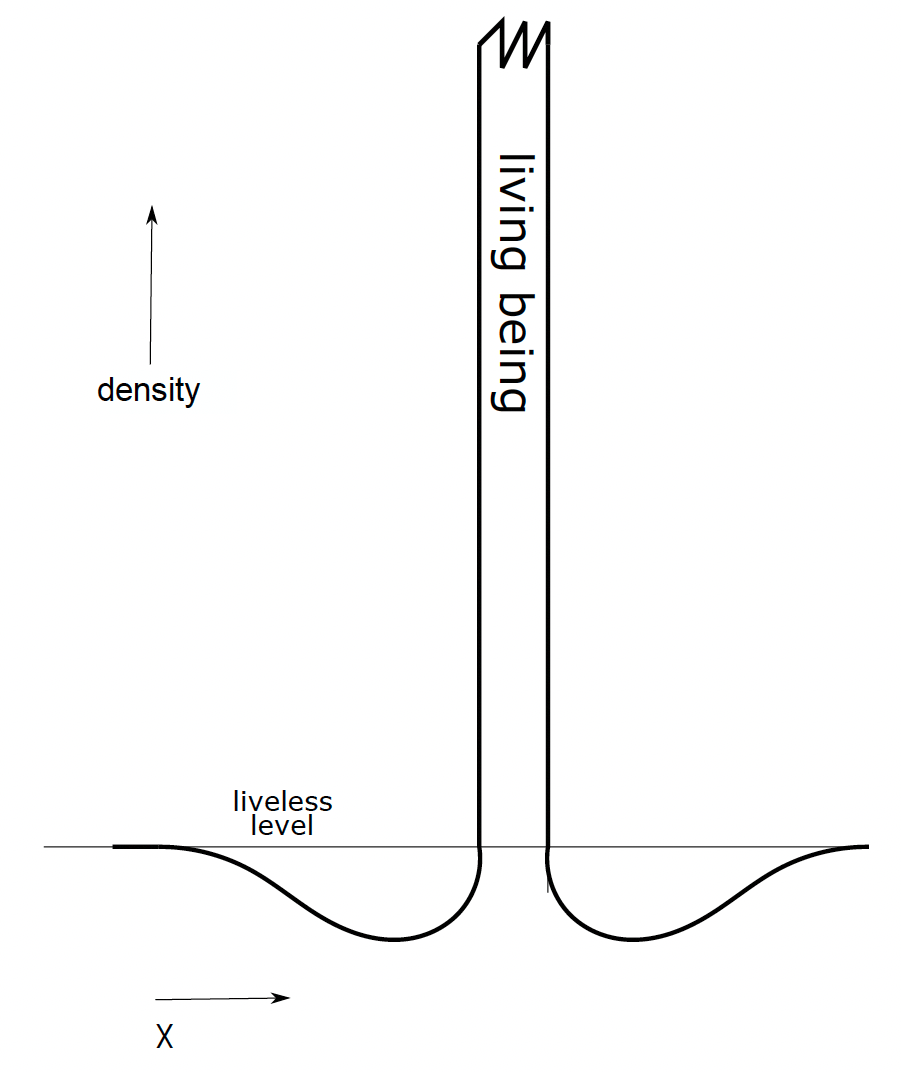}
      \caption{Vital density field distribution for a living being and its environment.
              }
         \label{fig1}
   \end{figure}

That is to say, a living being is not simply a highly dissipative system that consumes large quantities of energy in order to preserve its complexity, but also, and this must be emphasized, it ejects a great quantity of entropy below the lifeless density into its immediate surroundings. The density distribution is illustrated in Fig.~\ref{fig1}. A living organism absorbs energy from and \emph{entropizes} its surroundings.

To measure the density, we can benefit from Eq.~\ref{eq3} only when steady state conditions are satisfied. This can be achieved inside a living being or in an inhospitable medium, being nearly zero in an infertile medium. Inside a living being, the value of the densities would be as difficult to estimate as the specific entropy. The units of density would be the inverse of specific entropy, or entropy per baryon (i.e. K/erg, for example), or would be non-dimensional when the so-called system of geometrized units, with c, G and Boltzmann-s constant equal to unity, is adopted, as is usual in cosmology and other astrophysical studies. We consider life to be an astrophysical issue. 

\subsection{Vital field waves and vitons}

We define vitons as particles associated with vital waves in accordance with the interpretation of the wave-particle duality. We demonstrate, with the two basic vital field equations, the existence of waves in the vital density field. We now consider the external medium surrounding living organisms. Using the method of linear perturbations and introducing small changes, magnitudes susceptible to variation when traversed by the wave are transformed in the following way:

$$
 \begin{array}{l}
  \rho \rightarrow \rho + \rho' \\
  \vec{v} \rightarrow \vec{v} + \vec{v'}
 \end{array}
$$

Unprimed magnitudes correspond to the unperturbed medium, and primed magnitudes are the perturbed magnitudes when traversed by the waves. 

We introduce the following simplifications: (a) $\rho$ has no gradients and no temporal variations; (b) the perturbationless velocity is zero; (c) the perturbations are very small, so that terms implying a product of two perturbed terms are negligible; and (d) there are no specific entropy variations far removed from living beings, so that the term involving s does not give rise to further complications. The waves would exist even without these simplifying assumptions, but their mathematical treatment and physical behaviour would be less simple. In particular, the propagation of vitons inside living beings would be not so simple, where gradients of specific entropy due to different organs would render the wave equation more realistic.

From the equation of continuity we get
\begin{equation}
 {{\partial \rho'}\over{\partial t}} + \rho \nabla \cdot \vec{v'} = 0
  \label{eq7}
\end{equation}
and the equation of motion gives
\begin{equation}
   \rho {{\partial \vec{v'}} \over {\partial t}} + k_1 \nabla \rho' = 0
\label{eq8}
\end{equation}	
				
We derive the first with respect to time and find the divergence in the second, and, since the spatial and temporal derivatives are independent and therefore commute, we obtain directly:
\begin{equation}
   {{\partial^2 \rho'} \over {\partial t^2}} = k_1 \nabla^2 \rho'
   \label{eq8prima}
\end{equation}
which is the wave equation. Now, $k_1$ is the wave velocity squared; that is to say, the velocity of the vitons. We may suppose that $k_1 = c^2$, where $c$ is the speed of light, since these waves can propagate in vacuo. The wave velocity (or the velocity of the vitons) could be equal to the velocity of light or close to it. In the geometrized units, that are the custom in cosmology, with $c = 1$, $k_1 = 1$ also. 

We could likewise derive a wave equation in terms of velocity. The propagation of the waves is accompanied by small fluctuations in velocity, $\vec{v'}$. It is to be noticed that this velocity is not the wave velocity, equal to $k_1^{1/2}$. In a fluid also, the perturbed velocities by the passage of the wave are not equal to the sound speed.

Inside a living being, this wave equation can be produced in the following way. A mutation can be produced at any given moment, giving rise to an abrupt variation in entropy, which is equivalent to a sharp change in vital density.  A density jump in a medium with the capacity to support waves generates the wave. In other words, mutations can create a wave and the emission of vitons. The amplitude of this wave, being a purely geometric factor, diminishes as the inverse square of distance. Vitons always travel at the speed of light or at least at a very high speed. There is some similarity between (energy-photons) and (entropy-vitons).

As in the case of normal mutations, VID mutations can be produced by auto-replication or induced by external agents, X-rays, and other means. We cannot provide examples of VID mutations, as their effects are noticeable only over long time scales (consider, for example, the huge range of complexity of mammals).

\section{Chemistry of life}

The origin, evolution and fate of living beings are the result of the interrelation of both, physical and chemical processes. In this paper we focus on the physical ones but it is clear that chemistry, more precisely, biochemistry, is of the greatest importance to understand life, both terrestrial and extraterrestrial. The knowledge of biochemistry is at present incomplete but this is due to the great complexity of the problem rather than the chemistry fundaments themselves require to be revised. Biochemistry of cosmic life has been reviewed in several papers. The reader is addressed to two excellent books by \citet{schulze} and \citet{trigo}, with references therein. These authors not only examine the life chemistry in the Earth, but also speculate on alternative cosmic niches and study biochemistry under a cosmic perspective. 

It is evident that life can develop at least when physical-chemical conditions are available. A suitable habitat requires chemical heterogeneity. In Earth, life macronutrients are C, H, O, N, K, Ca, P, Mg and S to form prebiotic monomeric inorganic compounds such as H$_2$, N$_2$, CO$_2$, NH$_3$, HCN \dots Of great importance was the emergence of photosynthesis, but this is a complicated mechanism that could not be at work in the primitive forms of life. Early chemolithotrophic life was important using redox metabolic reactions, in particular the oxidation of hydrogen to water, with 2.6 eV per reaction, coupled with the reduction of CO$_2$ (metanogenesis, 1.4 eV per reaction), which is as much energetic than light (about 2 eV per photon depending on wavelength). Early cometary and meteoritic bombardment also delivered water and organic compounds thus enriching primordial chemistry as first proposed by \citet{oro}. Carbonaceous chondrites are rich in organic molecules, as witnesses of the protoplanetary disk chemistry.

Cyanobacteria and other chemoautotrophic organisms dramatically produced oxygen changing the atmosphere, rendering oxidation conditions. Life went from anaerobic to aerobic. This fact favoured the carbon chemistry with its ability to form complex stable molecules, starting with methane and other alkanes, reaching amino acids leading to proteins through connections by peptide bonds. Also, cycloalkane and polycyclic aromatic hydrocarbons (already present in the interstellar medium, in particular in molecular clouds) produced rings and later three-dimensional macromolecules with a non-racemic left-handed chirality. The carbon polymeric chemistry is able to build an almost unlimited range of molecules with strong covalent bonds, providing a very rich repertoire of metabolic reactions. This change of chemistry illustrates that not only life is adapted to the atmosphere but also that the atmosphere is adapted to life.

Life on Earth began with microscopic anaerobic prokaryotic cells, starting with LUCA (Last Universal Common Ancestor). Two billions later the emergence of eukaryotic cells took place as a fusion of two or more prokariotes \citep{margulis95}, with sizes grater by an order of magnitude. The emergence of multicellular life is a poorly understood step. About 541 millions years ago the Cambrian explosion was the origin of bilaterian and homeothermic phyla. This homeothermia was an important step as metabolic reactions adapted to narrow to optimal temperature ranges. On the other hand, homeothermia provided an efficient way to heat the environment thus expelling entropy for keaping its disequilibrium state.

We cannot present a brief overview of the very wide branch of cosmochemistry. We limit ourselves to brief comments on biochemistry topics closer to the physics proposed in this paper: a) the necessity of a solvent, b) the required boundary and c) the replication mechanism.

\noindent (a) The necessity of a solvent. Life is fundamentally a liquid based process. Terrestrial life uses water as a solvent especially efficient due to its high dipole moment. Atoms and molecules can move around nearly free to undergo chemical reactions. There is a perfect compatibility between carbon polimeric chemistry and water as a solvent. However, life was perhaps not originated in oceans or sees but in interfaces liquid-solid like in puddles alternating desiccation and rehydration, or porous media like pumice or even ice in contact with liquid water. A solid phase could preserve initial and later decreases of entropy.

\noindent (b) The required boundary. The existence of boundaries of living beings is of great importance to isolate the regions of disequilibrium inside from the environment closer to thermodynamic equilibrium, to isolate a region with very low entropy and a medium with very large entropy. Biomembranes preserve the high free energy state of the system from dissipation, encapsulating and confining a high concentration of interacting solutes and macromolecules. But isolation cannot be absolute. The semi-permeable boundaries must provide a selective transport of nutrients and waste. For this difficult task anphiphylic lipids (fatty acids, phospholipids) are essential as are hidrophylic toward the aqueous environment and hydrophobic toward the cell interior. The emergence of lipid protomembrane boundaries was a great jump in the early evolution of life.

\noindent (c) The replication mechanism. At present, in Earth\'s life, the information is stored in the one-dimensional form of a linear code (DNA) translated into the there-dimensional structure of proteins but, to reach this complex, precise, high fidelity mechanism, many previous insufficiently known steps were to be surmounted. There is at present a great difficulty to identify the first replicators. \citet{kauffman95,kauffman} proposed that the origin of life and the transmission of hereditary information was not due to a single molecule but to a network of interacting catalytic molecules. Probably there was a pre-RNA world followed by an RNA world before reaching the DNA-RNA world in which RNA is now an intermediate in the flow of information and responsible of the production of catalytic and autocatalytic encymes. There must be a transmission of information from parental to descendants or offs-springs to assure an autonomous maintenance of phyla, a great biochemical challenge at early epochs. These molecular codes permit the evaluation of the entropic content of living individuals.

Much work must still be done to understand the establishment of terrestrial life chemistry, and still much more work to speculate about extraterrestrial life adapted to other energy sources and other habitats, but we do not suggest that new chemistry fundaments are needed. On the other hand, it is here suggested that complementary physics as introduced in this paper, should favour the evolution to reach the chemical complexity of life, rendering some difficult steps more plausible.

\section{Results}

\subsection{Mutations driven by vital density}

Standard mutations, which occur completely at random, may give rise to both positive and negative variations in entropy in the molecular code. We add another kind of mutation driven by the vital density field and characterized solely by a diminution in entropy. We call these \emph{VID mutations} (Vital Driven mutations). The existence of VID mutations is one of our fundamental hypotheses. Both types are random, coexist, are subject to natural selection, and drive the evolution of a given species. Although standard mutations might well dominate, the introduced kind are also essential because they enable us to better understand the progressive appearance of complexity through time, a problem that we are using to test our model. 

The simplest mathematical way to express the effect of VID mutation is to suppose that
\begin{equation}
  ds = -k_3 \rho dt
  \label{eq9}
 \end{equation}
 
In other words, the temporal diminution in specific entropy is proportional to the vital density. Although the mutations are quantized in nature, after long periods of time a continuous description may be permitted, as expressed in Eq.~\ref{eq9}.

The real evolution of species is the result of the joint action of standard and VID mutations, together with multiple interruptions, bifurcations, etc. The graph of specific entropy as a function of time is extremely complex, as testified by the history of evolution. In order to analyse the isolated effect of VID mutations we plot the curve of the variation of entropy against time by considering the ideal effect where VID mutations are alone responsible for this variation. By isolating the tendency of VID mutations in this way we may insert the result into a global interpretation involving all evolutionary effects.

If we combine Eqs.~\ref{eq9} and ~\ref{eq3},
 \begin{equation}
   ds = - k_3 {k_1 \over k_2} {1 \over s} dt = -k_4 {1 \over s} dt
   \label{eq10}
 \end{equation}

This differential equation integrates easily to produce:
 \begin{equation}
   {s \over s_0} = \sqrt{1 - {{2 k_4} \over {s_0^2}}t}
   \label{eq11}
 \end{equation}
 
\begin{figure}
   \centering
   \includegraphics[width=\hsize]{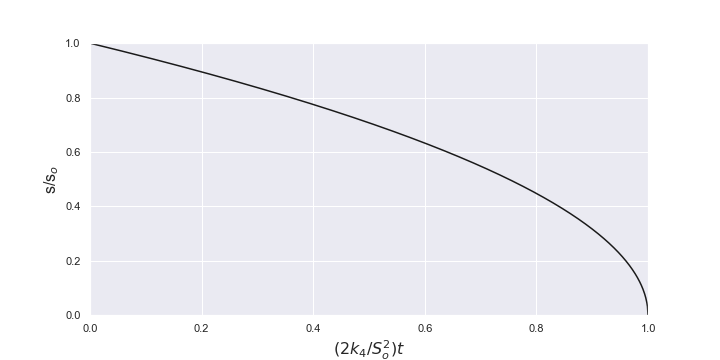}
      \caption{Variation of specific entropy with time due to the isolated effect of VID mutations.
              }
         \label{fig2}
   \end{figure}

We avoid a constant of integration by choosing a suitable time origin. The constant $s_0$ represents the value of specific entropy at infinity, approaching the lifeless level. The value of $k_4$ must be very small for the effects of VID mutations to be applicable on evolutionary time scales. After a time interval of, say, 106 yr, $k_4$ would be of the order of 3$\times$10$^{-14}$ s$^{-1}$. We calculate the value of the lifeless specific entropy per baryon to be of the order of $s_0 = 5\times 10^{-15} erg K^{-1}$. The estimation of this value is given in an appendix. Fig.~\ref{fig2} shows the variation curve.

We see from Eq.~\ref{eq11} that for low times the specific entropy decreases slowly until approximately the characteristic time $s_0^2 k_4^{-1}$. At larger times the decrease becomes faster until reaching its minimum value, which is zero as the entropy is positive. The value of zero corresponds to DNA as a pure crystal,  equivalent to zero temperature. It is therefore an ideal limit.

On small time scales the action of VID mutations will not be very great; when, however, it has attained a significant level of complexity, it could predominate. The specific entropy would eventually correspond to what we may term the \emph{zero-entropy species}, beyond which there could be neither greater complexity nor further evolution. This ideal state would not, however, be unchangeable, depending on ecological or environmental conditions, on the action of normal mutation, etc. It is to be emphasized that this plot considers only the isolated effect of VID mutations. 

The zero-entropy species is not the end of evolution, because normal mutations are always at work and can reverse the trend. Other evolution paths can then be explored. This ideal end will be reached only if only VID mutations were present, which is by no means assumed. In the real world other effects such as bifurcations and extinctions could take place. On the other hand, there are many paths to reducing entropy; there would be many end points. There would be not just one zero-entropy species, but many, and there is no way of telling which one of them would reach the final state chosen by natural selection. The most complex species need not be the best adapted ones, so that almost perfect, but fragile and ailing, species might result that are doomed to extinction from the start. 

If we put $s = 0$ in Eq.~\ref{eq11}, we get:
 \begin{equation}
   s_0^2 = 2 k_4 \tau
   \label{eq12}
 \end{equation}
where $\tau$ is the characteristic time for VID mutations to reach the condition of zero-entropy species. Bearing in mind the estimated values of $s_0$ (see Appendix) and assuming $\tau$ to be of order $10^9$ yr, somewhat less than, but of the order of, a Hubble time, we have an estimation of the value of $k_4$ of the order of $10^{-14} s^{-1}$.

It is tempting to identify the zero-entropy species as some kind of \emph{superior} species, even to the point of imagining \emph{superior humans}, but such a notion is untenable, given that a zero-entropy species would be the result of its evolutionary path, which cannot be determined beforehand, so a zero-entropy shark would be just as conceivable as a zero-entropy human or a zero-entropy extraterrestrial, for example. It is possible that the shark species has already reached a state close to zero entropy, further evolutionary progress being difficult, given its phylogenesis. There is, then, no anthropic principle at work in our hypothesis.

\subsection{Spreading of mutations}

\begin{figure}
   \centering
   \includegraphics[width=\hsize]{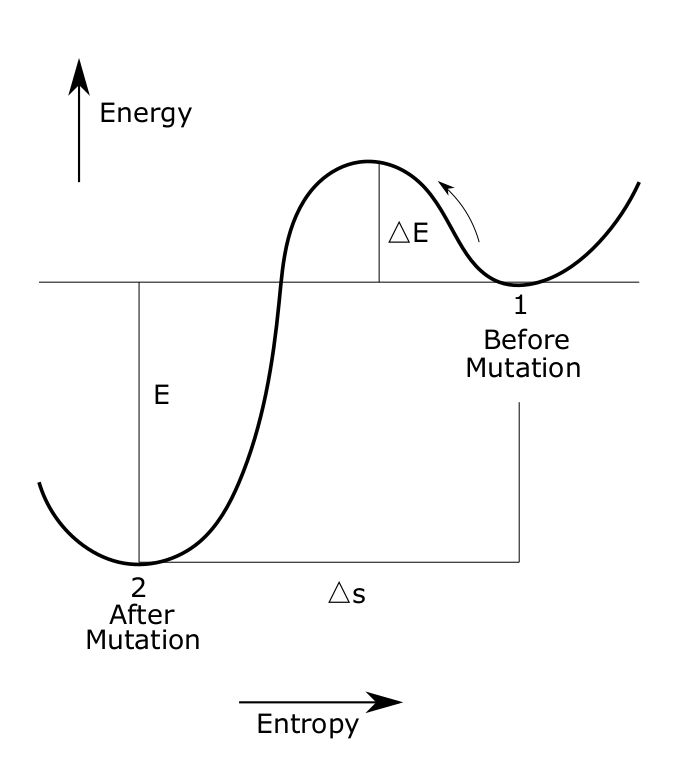}
      \caption{Energy-entropy diagram showing states before and after a VID mutation.
              }
         \label{fig3}
   \end{figure}

The great stability of many molecules, particularly DNA or any other equivalent code, implies that they are in a state of relative minimum energy. For a mutation to occur and a different state to be reached, a high potential energy barrier, $\Delta E$, must be surmounted, and that barrier must be much higher than the thermal energy, $kT$  ($k$ being Boltzmann'ss constant and $T$ the temperature), so the molecule does not easily change its state. This condition, equivalent to a very low temperature, was used by Schr\"odinger, who considered the gene to be an aperiodic solid, ideally at a temperature of absolute zero. Aside from a leap in entropy, $\Delta S$, the mutation would undergo, independently of $\Delta E$, a change in energy, $E$, from its initial to its final state, corresponding to a photon emitted in the mutation (see Fig.~\ref{fig3}).

The existence of vitons may clarify the manner in which a mutation (whether or not favourable, with $\Delta S$ either positive or negative) originating in an individual of a species can spread to other individuals of the same species to the point of mutating the entire species. Mutation is an entropic leap in a DNA molecule that creates a sharp increase in vital density, giving rise to the emission of a viton. When a viton meets a DNA molecule of the same individual, or of other individuals of the same species, it will provoke the same entropic leap, $\Delta S$, that gave rise to it, so that the viton will have propagated the mutation of the emitting DNA molecule to the receptor DNA molecule. In its turn, the receptor will emit another viton of exactly the same entropic leap, $\Delta S$, thereby producing a scattering effect in the density waves. An analogy may be drawn with the processes of emission and absorption of photons. This mechanism would be at work in DNA molecules as well as in any other molecule used as information code. 

For transmission to be effective the viton must carry information on both $\Delta S$ and $\Delta E$, the energetic leap required for the mutation to take place. The viton would not only announce that a mutation is about to be produced but also would supply the energy to make it happen. No problem arises if the mutational leap gains more energy than it loses, i.e. if $E$ is greater than $\Delta E$ (see Fig. \ref{fig3}). 

When the receptor organism in turn becomes an emitter, the mutation will be transmitted to neighbouring beings belonging to the species. In this way a mutation originating in an individual may become a mutation of the entire species.

A quantum leap in entropy could correspond numerically to more than one mutation. The receptor organism might undergo a different mutation from that of the emitting organism. This error in transmission does not imply any evolutionary disadvantage since, for progressive mutation to occur, diminution in entropy is more important than precisely where the dislocation in the initial DNA takes place.
 
Standard mutations and VID mutations are not excluding each other. However they are independent as they are produced by different mechanisms. Standard mutations are the results of random changes in the DNA molecule (excluding those produced by X-rays...) VID mutations are induced by vital waves. Let us clarify this point by finding other similar effects in classical systems. In a hot metal, for instance, electrons may escape the metal as a result of thermal noise. In addition, if the metal is exposed to UV radiation the electrons also escape due to the photoelectric effect. The two mechanisms co-exist, both contribute to the escape of electrons and are independent. The UV photons induce another type of ejecting electrons.

\section{Discussion}

With regard to the problem of life origin and evolution there are three types of contributions: an astrophysical phase concerning the production of rather complex biomolecules and the provision of habitable environments, the contribution of genetics to identifying how evolution proceeds, and the contribution of astrophysics in dealing with entropy and the probabilities involved in evolution. Our paper only deals with the third aspect. How biomolecules are formed in space, in the proto-solar system and in atmospheres is of course crucial but are beyond the scope of our paper. Here we restrict ourselves to the entropic problem of life.

When and how the vital field is originated in the Universe? Our vital field is present in all coins of the Universe, even with in general very low values of density, and its integrated value is a constant in time. The origin itself should be a cosmological problem that is far from the present preliminary state of the theory. 

The existence of a vital field in the Universe provides an effective scenario for other forms of life in the Universe and sheds new light on a number of phenomena that are still controversial in the physics of life and its evolution. With this model we have been able to explain two of these phenomena: the appearance of species of growing complexity in Earth's history. This growing complexity should be a basic property also in other environments in the Universe. We have also explained the rapid dissemination of a mutation in an individual of the species that culminates in the mutation of the entire species. 

This theory could be generalized to incorporate relativistic and quantum aspects, but given the exploratory and novel focus of this study, we have preferred to introduce it in a simple mathematical way.

We have considered the special case of living beings, but the vital field can be present in any place in the Universe, even in liveless points far from living entities. Similarly, vital waves and vitons may be present anywhere.

The theory proposed here makes reference to the genome, but is quite independent of the specific mutation mechanism. When considering physical laws, which are in principle applicable to the entire Universe, these ideas could have general implications concerning the emergence and evolution of life as a cosmological fact.

Mutations, both standard and those driven by the vital field, must accept the verdict of natural selection. For standard mutations $S$ may be positive or negative; for VID mutations $S$ will be negative. The $\left[s,t\right]$ curve for VID mutations has a gentle slope for short times, but the slope becomes more pronounced for long periods of time and low specific entropy. This implies that, if the evolution reach a very low level of entropy, VID mutations could become of greater importance. To put it succinctly, complexity breeds complexity.

Isolated VID mutations would lead to ideal zero-entropy species, but VID mutations, still with the restriction that $\Delta S <$ 0, are also aleatory. Furthermore, such species are the result of their evolutionary path. For that reason we say that our theory is not teleological. Neither does man possess any special privilege in our model, since it is not based in any way on the anthropic principle. 

The concept of \emph{zero-entropy species} is an idealization and would correspond to a minimum of entropy depending on the evolution path. A rabbit can be close to a \emph{zero-entropy rabbit}, for example. In this model, there is no end point or point toward which it evolution converges; on the contrary, there are as many end points as there are ways of reducing entropy at every step. As stated in Sect. 3, the final end point will be the result of natural selection. Randomness here is limited but not eliminated. We just reduce the possibilities for VID mutations and preserve standard mutations. The adaption to changing environmental and ecological conditions is here considered as fundamental, too. 

The density of the vital field tends to reach a distribution with very high values inside a system of very low specific entropy. A living being is conceived here as a system of very low entropy animated by very high values of the density of the vital fluid at the same place occupied by the system. However, we have not focused this study on the concept of life itself, but rather on its evolution.

We have not considered sexual reproduction because the theory could be applicable to any form of life in the Universe for which we do not know the reproduction mechanism. In any case we do not at all exclude the advantages of sex to the rise of new species and biodiversity.

The densities in the living being contribute to an evolution that favours the diminution of entropy and, consequently, an increase in density. There is an intrinsic instability: low entropy produces increase of density (Eq.~\ref{eq1}) and density reduces the entropy (Eq.~\ref{eq9}). The fist process is fast and the second slow. Although this instability takes place on evolutionary time scales, it is present in every phase. 

It is even present at the origin of life. In effect, inert matter can undergo natural fluctuations from thermodynamic equilibrium; these lead to fluctuations in the distribution of specific entropy, which imply vital density fluctuations, which in turn produce a prezoic type of heterogeneity that, because of this instability, leads to the first prokaryotes. In terms of this model, the generation of life would not need to be stimulated by mechanisms different from those of evolution, although the generative process itself would require one of its most significant leaps.

One could argue that entropy is a magnitude that was introduced with other purposes in mind, and that the notion of a quantum of entropy does not fit into the original conception of entropy. However, Schr\"odinger has already spoken of a quantum theory of life. In physics it has often been the case that magnitudes have successfully been given a much broader interpretation than when they were first introduced. Also, the concept of energy, although both familiar and abstract, only began to be appreciated in the nineteenth and twentieth centuries, more or less at the same time in which the concept of entropy was established. On the other hand, we only consider quanta of entropy when dealing with mutational transitions.

\section{Conclusions}

The ideas presented here are not an alternative to the standard evolution paradigm, but rather introduce novel physical concepts that speed up the evolution of life, thus rendering more probable the growth of complexity.

There is an intrinsic instability: low entropy produces en increase of vital density; vital density produces a decrease of entropy. The first mechanism is very fast and the second very slow, with typical characteristic evolution times. As a result of this instability, the distribution of both low entropy and vital density tends to reach very high values in small places, i.e. in living beings. 

The higher the degree of complexity the faster becomes the decrement in entropy; therefore the faster the increase of vital density, the faster the increase of complexity. The net effect of the VID mutations introduced here is that complexity breeds complexity.

The similarity between the Eulerian fluid and the vital fluid considered here, enables the existence of vital waves and their corresponding particles (vitons). The speed of vitons should be vey high, as great as the light speed or slightly lower. This provides another way of accelerating evolution in terms of the faster disemination of mutations among individuals of the same species. Energy-photons and entropy-vitons are another parallel that could be emphasized. 

When we view the complexity of life from a new perspective, specific properties and laws emerge that we have tried here to identify. We have seen how simple physical propositions, together with straightforward and simple mathematics, have sufficed to clarify some of the most controversial questions in the evolution of life on Earth and may even be used to establish general guidelines for understanding life in the Universe. The growth of complexity could be an intrinsic characteristic of any kind of life in diverse astrophysical environments.

\begin{acknowledgements}
\noindent Acknowledgements: We are very grateful to an anonymous referee for his/her valuable comments. We are also very grateful to Estrella Florido and Terry Mahoney. Helpful discussions with our
colleagues Jos\'e I. Illana, Antonio Mampaso, Manuel Masip, Evencio Mediavilla, Fernando
Moreno, Enrique P\'erez, Ram\'on Rom\'an and Ignacio Trujillo have improved the paper.  
\end{acknowledgements}

%
%



\begin{appendix} 
\section{Estimation of the value of the maximum specific entropy, $s_0$}
Let us consider Boltzmann's equation relating to entropy and thermodynamic weight (the number of microstates compatible with a macrostate):
\begin{equation}
S=k \ln{W}
\end{equation}

We assume that the inert matter is hydrogen. In one gram there are $N_a$ hydrogen atoms, $N_a$ being Avogadro's number (approximately $6\times 10^{23}$). We can exchange each atom by each atom and always obtain the same macroscopic result. The value of $W$ in the inanimate universe will then be ($6 \times 10^{23}$)!, where the symbol '!' means factorial. This seems difficult to calculate but we can use Stirling' formula for very large numbers:
\begin{equation}
  \ln{n!} = n \ln{n} - n
\end{equation}

In our case
\begin{equation}
\ln{6\times 10^{23} !}=54.7\times6\times 10^{23}-6\times 10^{23} \simeq 3\times 10^{25}
\end{equation}

The entropy of one gram in the inanimate universe will be of the order of 
\begin{equation}
1.38\times 10^{-16} \times 3 \times 10^{25}=4\times 10^9  erg/K
\end{equation}
and the entropy per baryon:
\begin{equation}
s_0=5\times 10^{-15}  erg/K
\end{equation}

In cosmological calculations 'geometrical units' are often used with $c$, $G$, $k$ = 1; the specific entropy then becomes non-dimension and has the value $s_0 = 30$. With this value in Eq.~\ref{eq12} and assuming a characteristic time of evolution of $10^9$ years we obtain $k_4$ to be of the order of $10^{-14} s^{-1}$.

\end{appendix}

\end{document}